\documentclass[acmlarge, nonacm]{acmart}

\AtBeginDocument{%
  \providecommand\BibTeX{{%
    \normalfont B\kern-0.5em{\scshape i\kern-0.25em b}\kern-0.8em\TeX}}}

\usepackage{multirow}
\usepackage{booktabs}
\usepackage{ifthen}
\usepackage{color}
\usepackage{url}
\usepackage{xspace}
\usepackage[T1]{fontenc}
\usepackage{enumerate}
\usepackage[linesnumbered,ruled,vlined]{algorithm2e}
\usepackage{array,multirow,graphicx}
\usepackage{float}
\usepackage{balance}
\usepackage{tikz}
\usepackage{calc}
\usepackage{subfigure}
\usepackage{listings}
\usepackage{url}
\usepackage{xspace}
\usepackage{hyperref,endnotes}
\usepackage{xcolor}
\usepackage{array}
\usepackage{multirow,makecell}
\usepackage[misc]{ifsym}
\usepackage{framed}

\newcommand{\para}[1]{\smallskip\noindent{\bf {#1}. }}
\setcopyright{acmcopyright}
\copyrightyear{2021}
\acmYear{2021}

\renewcommand\footnotetextcopyrightpermission[1]{} %

\begin{document}

\title{Demystifying Removed Apps in iOS App Store}

\author{Fuqi Lin}
\email{linfuqi@pku.edu.cn}
\affiliation{%
  \institution{Peking University}
  \streetaddress{No.5 Yiheyuan Road}
  \city{Beijing}
  \country{China}
  \postcode{100871}
}

\makeatletter
\let\@authorsaddresses\@empty
\makeatother

\begin{abstract}
  With the popularity of mobile devices, mobile applications have become an essential part of people's lives. To provide secure mobile application download channels for users, various modern app markets are maintained by different companies. For example, Google maintains Google Play for Android users, while Apple maintains App Store for iOS, iPadOS, and MacOS users. Though app markets have come up with strict policies which impose restrictions on developers to avoid the potential harmful applications, we still have quite limited knowledge on the process of app vetting and the status of potential harmful apps. To fill this gap, this paper takes the initiative to conduct a large-scale and longitudinal study of removed apps in the iOS app store. Our analysis reveals that although most of the removed apps are low-quality apps, a number of them are quite popular. Furthermore, the mis-behaviors of these apps are reflected on app metadata, which makes it possible to distinguish potential harmful apps.
\end{abstract}

\maketitle

\section{Introduction}
\label{sec:introduction}

Mobile apps have seen widespread adoption in recent years. As of 2nd quarter 2020, there are over 2.7 million apps available in Google Play and over 1.8 million apps available in iOS app store~\cite{AppNumber}. Although the mobile app ecosystem has seen explosive growth, \textit{app quality remains a major issue across app markets}. A number of low-quality apps, fake/cloned apps, fraudulent apps, illicit apps, Potentially Harmful Apps (PHAs), and even malware were recurrently found in Google Play, iOS app store and third-party app markets~\cite{wang2018beyond,iOSCopycat,iOSMalware,wang2019understanding}, posing great threats to mobile users.

To improve app quality and nip the potential threats in the bud, most app markets including Google Play and iOS app store have released strict developer policies, along with inspection and vetting processes before app publishing. For example, Google Play has released a set of developer policies~\cite{GoogleDeveloperPolicy} that cover 10 main categories, and iOS app store has published a detailed review guidelines~\cite{Apple20} covering five main parts including safety, performance, business, design and legal. 
Ideally, mobile apps that break these policies should not be published on app markets.

However, it is challenging to automatically identify low-quality and potentially harmful apps in the app vetting process. There has been growing evidence~\cite{iOSXcodeGhost,iOSMalware-1,iOSMalware-2,iOSAdware} showing the ineffectiveness of app vetting in both Google Play and iOS app store, i.e., a large portion of low-quality apps and PHAs sneak to app markets from time to time. Even for iOS app store, one of the most secure app markets due to its locked-down ecosystem, malicious apps~\cite{iOSMalware-1,iOSMalware-2} and aggressive adware~\cite{iOSAdware} were recurrently found. For example, it was discovered~\cite{iOSXcodeGhost} that hundreds of popular iOS apps were infected by XcodeGhost, a malicious Xcode development tool, affecting millions of mobile users directly.
Furthermore, some malicious and fraudulent behaviors cannot be explicitly identified during app vetting process. For example, previous work~\cite{AppPromotion-1,AppPromotion-2} suggested that apps can manipulate its metadata and generate fake reviews to boost its app ranking and search ranking. Some sophisticated apps (especially the HTML container apps) can behave normally during the app vetting process, while showing their illicit or malicious behaviors once getting in the market. 
These kinds of fraudulent behaviors, however, cannot be explicitly perceived based on binary-level analysis before they sneak into the market.

\textbf{Removed Apps.}
As a result, app markets have been removing low-quality and policy-violation apps continuously in a reactive manner, after the apps were listing in app markets for a certain time. For example, the number of apps in iOS app store peaked at around 2.2 million in 2017, but declined over the next few years (around 1.8 million apps in 2020), as Apple continues to remove low-quality apps and apps that break guidelines~\cite{AppleAppRemove-1,AppleAppRemove-2}. Although a number of reports and news media have mentioned removed apps, \textit{our research community still lacks the comprehensive understanding of the landscape of this kind of apps and the app maintenance (i.e., app removal) behaviors of app markets}. There remain a number of unexplored questions, e.g., \textit{how many apps were removed from the app market?} \textit{were they removed periodically or occasionally?} \textit{what are the practical reasons behind those app removals?}

\textbf{This Work.} In this paper, we present a large-scale longitudinal study of removed apps. To be specific, we focus on iOS app store, the official and largest app market for iOS apps. We first make great efforts to collect daily snapshot of the whole iOS app market (i.e., information of all available apps), with over 500 complete snapshots in total (from Jan 2019 to April 2020). 
By comparing the two consecutive snapshots, we can pinpoint which apps were removed, and their accurate removal date (\textbf{see Section~\ref{sec:studydesign}}). \textit{To the best of our knowledge, this is the largest and most comprehensive dataset of removed apps studied in the  research community}. This enables us to perform fine-grained analysis of removed apps.
We then characterize the overall landscape of removed apps (\textbf{see Section~\ref{sec:general}}), including the daily app maintenance behaviors across the span of 1.5 years, the life-cycle of the removed apps, their correlation with other app features (e.g., app popularity), and the developers of the removed apps. Our investigation suggests that although most of the removed apps are low-quality apps, a number of them are quite popular. It further motivates us to investigate the practical reasons behind the removed popular apps (\textbf{see Section~\ref{sec:removedpopular}}). We observe that a number of features from app meta information (e.g., app description, app comments, and app search optimization keywords) can be used to differentiate these removed popular apps effectively.
Therefore, we finally devise a practical app removal prediction model (\textbf{see Section~\ref{sec:detecting}}) based on machine learning techniques to identifying suspicious apps that should be removed.  

To the best of our knowledge, this is the first comprehensive study of app removal practice in iOS app store at \textit{scale}, \textit{longitudinally}, and \textit{across various dimensions}. In addition, compared to preliminary work of removed apps in Google Play~\cite{wang2018android} that are based on rather coarse-grained snapshots, this work comprehensively measured the overall landscape of removed apps  including the daily trend, app popularity, and app life-cycle, etc, and derived various insightful findings. Among  interesting results and observations, the following are prominent:
\begin{itemize}
    \item \textit{The number of removed apps is surprisingly higher than our expectation.} During the span of 1.5 years we studied, over 1 million apps were removed from iOS app market. Interestingly, app removal in iOS app store shows cyclical patterns. 
    \item \textit{Some developers have the tendency to release policy-violation apps.}
    Surprisingly, 73.45\% of removed app developers have all their released apps being removed. Majority of the removed apps are dominated by a limited number of developers.
    \item \textit{Removed ``popular'' apps are prevalent.} Although most of the removed apps are considered to be low-quality apps, 5\% of them are popular apps with high rank in app store and have gained a considerable number of user ratings.
    \item \textit{The removed ``popular'' apps can be identified proactively without analyzing their app binaries}. The features extracted from the app metadata can based used as strong indicators to differentiate the removed apps from normal ones.
    \item \textit{The removed apps can be identified prior to its actual removal date due to the initial signals it released.} By analyzing the evolution of its meta information only, we can pinpoint the suspicious policy-violation apps with roughly 80\% accuracy.
\end{itemize}

Our results motivate the need for more research efforts to illuminate the widely unexplored removed apps. We believe that our research efforts can positively contribute to bring user and developer awareness, attract the focus of the research community and regulators, and promote best operational practices across app market operators. To facilitate further study along this direction, we will release our dataset along with all the experiment results, to the research community (link removed due to anonymous submission).
\section{Background}
\label{sec:background}

\subsection{iOS App Store Guidelines}
\label{subsec:guidelines}

To improve app quality and eliminate potential security issues, Apple requires all iOS apps to go through a vetting process to determine whether they are reliable and perform as expected. The app vetting process adopts manual efforts with automated tools. 
App reviewers compare the app with Apple's public App Store guidelines\footnote{https://developer.apple.com/app-store/review/guidelines/}, making sure it does not show any undesired behaviors. 
On average, 50\% of apps are reviewed in 24 hours and over 90\% are reviewed in 48 hours\cite{AppReview}. If the submission is incomplete, review time may be further delayed or the app may be rejected. 
In general, the App Store Review Guidelines can be summarized as follows:

\textbf{Safety}. Apps should not include objectionable content that is offensive, insensitive, upsetting, intended to disgust, in exceptionally poor taste, or just plain creepy. 
Apps should not behave in any way that risks physical harm. Apps should implement appropriate security measures to ensure proper handling of user information.
    
\textbf{Performance}. Apps should be tested on-device for bugs and stability before submitting, and include demo account info. The metadata including app description, screenshots, and previews must be accurate and be kept up-to-date. Apps should satisfy the condition on both hardware compatibility and software requirements.

\textbf{Business}.
Apps that attempt to manipulate reviews, inflate chart rankings with paid, incentivized, filtered, or fake feedback, and expensive apps that try to cheat users with irrationally high prices will be hit hard and punished.  Ads displayed in an app must be appropriate and be limited to the main app executable. Besides, the payments of the app shall not be irregular.

\textbf{Design}.
Apps' design must meet the minimum standards for approval including no copycats, no spam, containing features, content, and UI, and hosting or containing compliant extensions, etc. Apps that stop working or offer a degraded experience may be removed from the App Store at any time.

\textbf{Legal}
Apps must comply with all legal requirements in any location where the app is available. It is paramount in Apple ecosystem to protect user privacy including data collection and storage, data use and sharing, and location services, etc. Intellectual property should not be infringed. Gambling, gaming, lotteries and VPN apps must not violate local laws.

\subsection{App Removal}

Although Apple has made great effort to improve the app vetting process, low-quality and policy-violation apps were found in the market from time to time. App removal is a common practice in the market.
Apps can be removed by either app developer or the app market. For the apps removed by app developer proactively, if the app binary has not been modified, the app developer can re-launch the app at any time, without going through the app inspection and vetting process. In contrast, for the apps removed by app market due to policy-violation reasons, the app developers should address all the issues and then submit for the app inspection and vetting process.
However, it is not feasible for us to infer whether the app was removed by app market.
Besides the apps that break review guidelines, mobile apps can also be removed by iOS app store due to political reasons (e.g., censorship by governments).

\section{Study Design}
\label{sec:studydesign}

We present the details of our characterization study on removed apps in this section. We first describe our research questions (RQs), and then present the dataset used for our study. 

\begin{table}[t]
\caption{An overview of the detailed information collected for the removed apps.}
\label{table:data-description}
\resizebox{0.8\linewidth}{!}{

\begin{tabular}{|c|l|l|}
\hline
\multicolumn{1}{|l|}{Type}                        & Data Field     & Description\\ \hline
\multirow{10}{*}{Objective}                       & App Name       & The name of the app \\ \cline{2-3} 
                                                  & App ID         & The ID of the app given by the app store\\ \cline{2-3} 
                                                  & Developer Name & The name of the developer of the app \\ \cline{2-3} 
                                                  & App Category   & The category of the app in app store\\ \cline{2-3} 
                                                  & Price          & The price of the app\\ \cline{2-3} 
                                                  & Status         &  Online, Removed or Relaunched\\ \cline{2-3} 
                                                  & Release Date   & The release date of the app\\ \cline{2-3} 
                                                  & Update Date    & The latest update date of the app\\ \cline{2-3} 
                                                  & Offline Date   & The date when the app is removed\\ \cline{2-3} 
                                                  & Relaunch Date  & The date when the app is relaunched (if existed)\\ \hline
\multicolumn{1}{|l|}{\multirow{2}{*}{Subjective}} & App Ratings    & The number of \{1,2,3,4,5\} ratings of the app   \\ \cline{2-3} 
\multicolumn{1}{|l|}{}                            & App Reviews    & Detailed app reviews from app users\\ \hline
\multicolumn{1}{|l|}{\multirow{2}{*}{Popularity}} & App Ranking    & App daily ranking in its category\\ \cline{2-3} 
\multicolumn{1}{|l|}{}                            & ASO Keywords   & The daily ASO keywords covered by the app  \\ \hline
\end{tabular}
}
\end{table}

\subsection{Research Question}

Our study is driven by the following research questions:

\begin{itemize}
    \item[RQ1] \textit{How many apps were removed from iOS App Store? What is the distribution of removed apps across time (e.g., daily) and space (e.g., category and app popularity) dimensions?} 
    Although a number of reports and news media mentioned app removal from iOS app store from time to time, our research community still lacks the understanding of the overall landscape. We are unaware of the app maintenance behaviors in app markets and the characteristics of the removed apps.
    \item[RQ2] \textit{Why were they removed from iOS App Store?} Although app markets have released strict developer policies, along with inspection and vetting processes before app publishing, a number of low-quality and potentially harmful apps were found in app markets after they were released, which then were removed after receiving massive user complaints. Understanding the reasons of app removal can help us understand the weakness of app vetting processes in iOS app store.
    \item[RQ3] \textit{Can we identify the removed apps in advance?} Existing efforts of app markets usually rely on reactive methods to identify potentially harmful apps after they sneak into the market, e.g., based on mobile users' feedback or the anomalies/damages caused by apps. Proactively identifying the removed apps before they are widely spread can help eliminate the potential risks they exposed. 
\end{itemize}

\begin{figure*}[htb]
    \centering
    \begin{center}
        \includegraphics[width=1\linewidth]{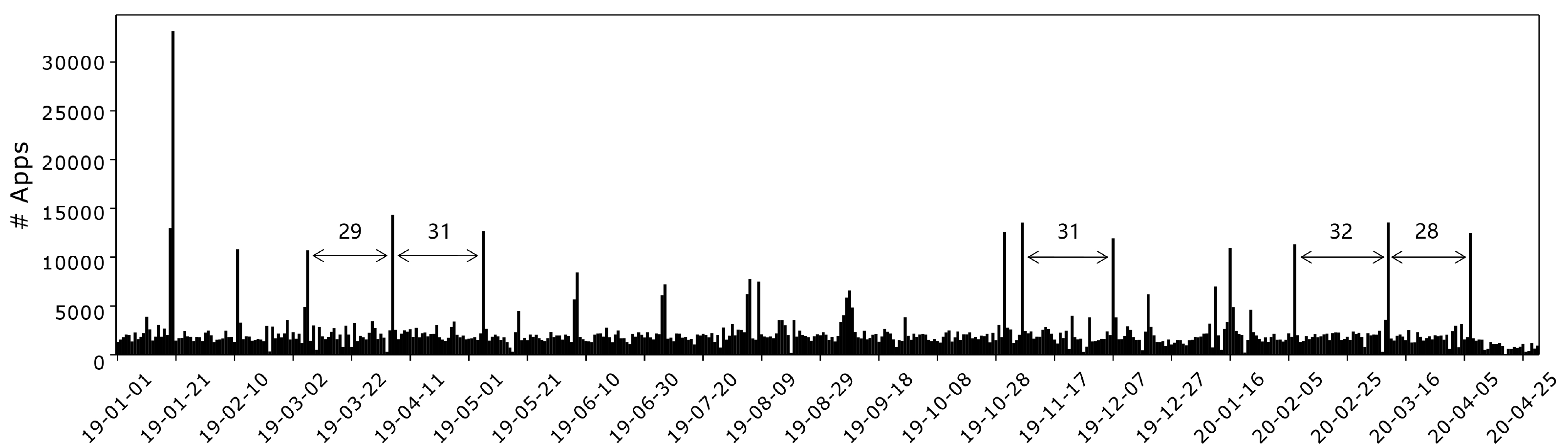}
        \caption{The number of daily removed apps in iOS app store (from January 1st, 2019 to April 30th, 2020).}
        \label{fig:daily-removed}
    \end{center}
\end{figure*}

\begin{figure*}[htb]
    \centering
    \begin{center}
        \includegraphics[width=1\linewidth]{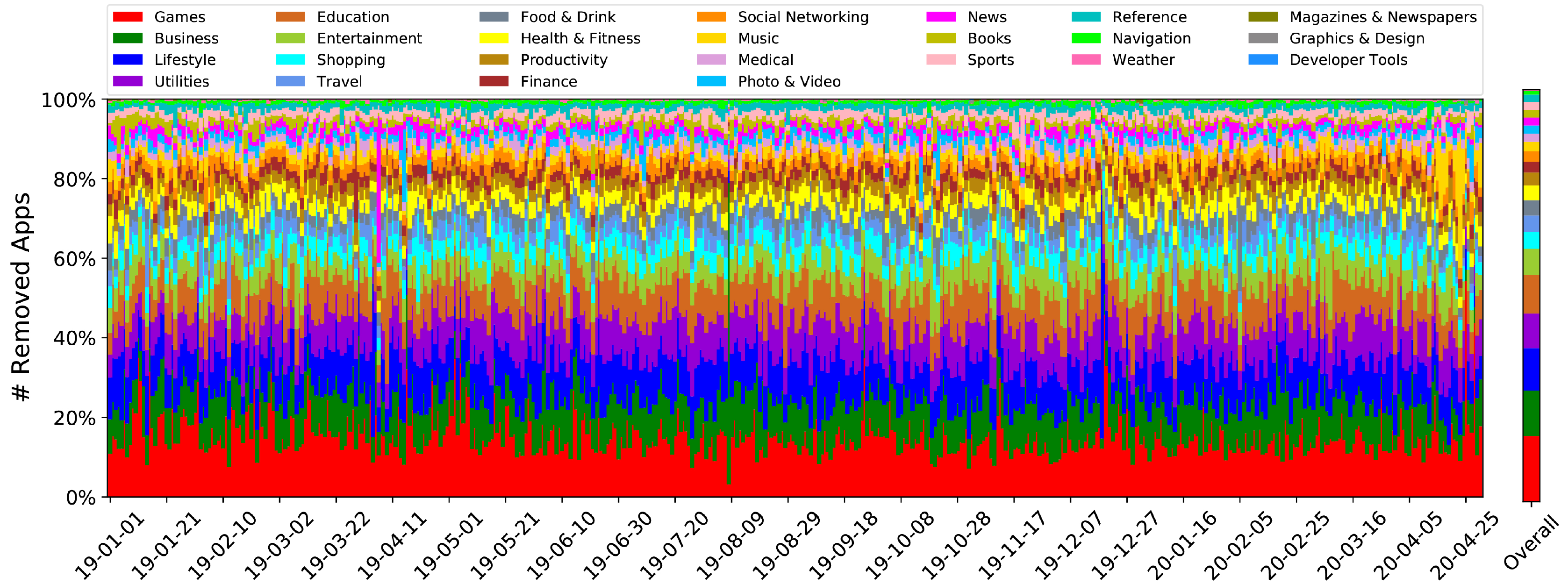}
        \caption{The proportion of daily removed apps across the 25 categories.}
        \label{fig:daily-removed-category}
    \end{center}
\end{figure*}

\subsection{Dataset Collection}\label{sec:dataset}

To answer the aforementioned research questions, we first need to harvest a comprehensive dataset of removed apps. However, it is non-trivial to compile such a dataset, as it requires great efforts to monitor apps in the app market continuously to check whether they were taken down, especially when we want to know their accurate removal date.
To this end, we cooperate with an anonymous leading mobile app intelligence company to get the daily removed apps from iOS app store.
The key idea of identifying removed apps is straightforward, i.e., we crawl the iOS app market to maintain a snapshot of the whole market everyday. By comparing the two consecutive snapshots, it is easy to know which apps were removed, and their accurate removal date.

We have collected the daily removed apps in iOS app store for 1.5 year, \textit{i.e.}, from 1st January, 2019 to 30th April, 2020. To the best of our knowledge, this is the only available and largest dataset on removed iOS apps in our research community.
Overall, there are 1,129,615 app removal records, which corresponds to 1,033,488 different mobile apps\footnote{Note that one app can be removed from the market more than once, see Section~\ref{subsec:lifecycle}.}.
To enable further analysis, we also collect the detailed information of the removed apps.
As shown in Table~\ref{table:data-description}, the information can be classified into the following three categories:
\begin{itemize}
    \item[1)] \textit{App Objective Information.} App objective information represents the data that created/defined by app developer/app market, e.g., app name, app ID, developer name, app price, etc. We have maintained a list of 10 objective information for each removed app. 
    \item[2)] \textit{App Subjective Quality.} 
    This kind of data is extracted from the information provided by mobile users. App markets provide multiple ways for users to leave feedback, which eventually are summarized to app ratings and detailed app reviews.
    \item[3)] \textit{App Popularity Information.} App popularity information can be reflected as its ranking in the market, and its app search optimization (ASO) capability. ASO~\cite{AppleASO} is the process of improving the visibility of a mobile app. The key idea is that, by optimizing the keywords (e.g., popular searching words) in their app metadata (e.g., name and description), their apps would appear popular (highly ranked) in the search results. App markets including Google Play and iOS app store asked the developers to avoid user testimonials, excessive details, misleading references to other apps and repetitive, excessive or irrelevant keywords. This kind of information was obtained from the cooperated app intelligence company by daily monitoring the search results of massive keywords in iOS app store. It will be used in Section~\ref{sec:removedpopular} and Section~\ref{sec:detecting} for characterizing removed apps with ranking fraud.
\end{itemize}

\section{Overview of Removed Apps}
\label{sec:general}

In this section, we seek to investigate the practical practices of iOS app market in its app maintenance behaviors. To be specific, we present the overall landscape of removed apps in iOS app store, including the number of daily removed apps across the span of 1.5 years, the life-cycle of the removed apps, their correlation with other app meta information, and the developers of them.

\subsection{Overall trend of removed apps}

\subsubsection{Overall Statistics}
Figure~\ref{fig:daily-removed} presents the number of daily removed apps in iOS app store. 
On average, 2,324 apps are removed daily. 
It can be observed that the number of the daily removed apps fluctuates periodically. 
For example, the peak on 20th January, 2019 indicates a large-scale app removal, with 33,173 apps removed. 
Interestingly, we can observe a number of peaks during the evolution, while the median of time interval between the adjacent peaks is 28.5 days (examples shown in Figure~\ref{fig:daily-removed}).
Thus, although apps can be removed daily from time to time, the large-scale app removal happens regularly every month in iOS app store.
\textit{We speculate that the app maintenance behaviors in iOS app store is cyclical (e.g., monthly), which may also be exploited by malicious developers}, i.e., using a number of spamming techniques~\cite{AppPromotion-1,AppPromotion-2} to reach to unsuspecting users during the ``silent time''.

\subsubsection{Category Distribution.}
Figure~\ref{fig:daily-removed-category} presents the proportion of daily removed apps across each category. 
It can be observed that apps in \texttt{Game}, \texttt{Business}, \texttt{Lifestyle}, \texttt{Utilities}, and \texttt{Education} categories account for most (over 50\%) of the daily removed apps.
This is consistent with the overall distribution of removed apps in each category (see the rightmost column in Figure~\ref{fig:daily-removed-category}). 
However, we can observe that the proportion of removed apps in some categories are significantly increased in certain days. For example, there is a large-scale app removal in \texttt{News} category on April 6th, 2019, which accounts for 36.48\% of all removed apps that day (compared with roughly 1.8\% in other days). 
Thus, \textit{we speculate that iOS app store would focus on certain kinds of apps in a period of time and remove the inappropriate ones intensively, which will lead to a large-scale app removal in the corresponding categories}. Some anecdotal evidence on news media~\cite{RemoveEvidence,RemoveEvidence2} supports our speculation.

\subsection{Popularity of Removed Apps}
We next investigate the popularity of removed apps. Generally, we expect that the removed apps are low-quality apps with few users. Note that, unlike Google Play that provides the informative download statistics of the apps, in iOS app store, we can only measure the popularity of apps based on their app ranking and the number of user comments~\cite{iOSStoreStudy}.

\subsubsection{App Ranking}
iOS app store provides the app ranking of top-1500 apps across different categories based on their real time popularity. Thus, during our dataset collection, we make efforts to harvest the daily app ranking information across the 25 categories. Without loss of generality, we define the \texttt{popular app} as the app that has ever been ranked in top-1500 in each category in this paper.

\begin{figure}
  \centering
  \includegraphics[width=0.8\textwidth]{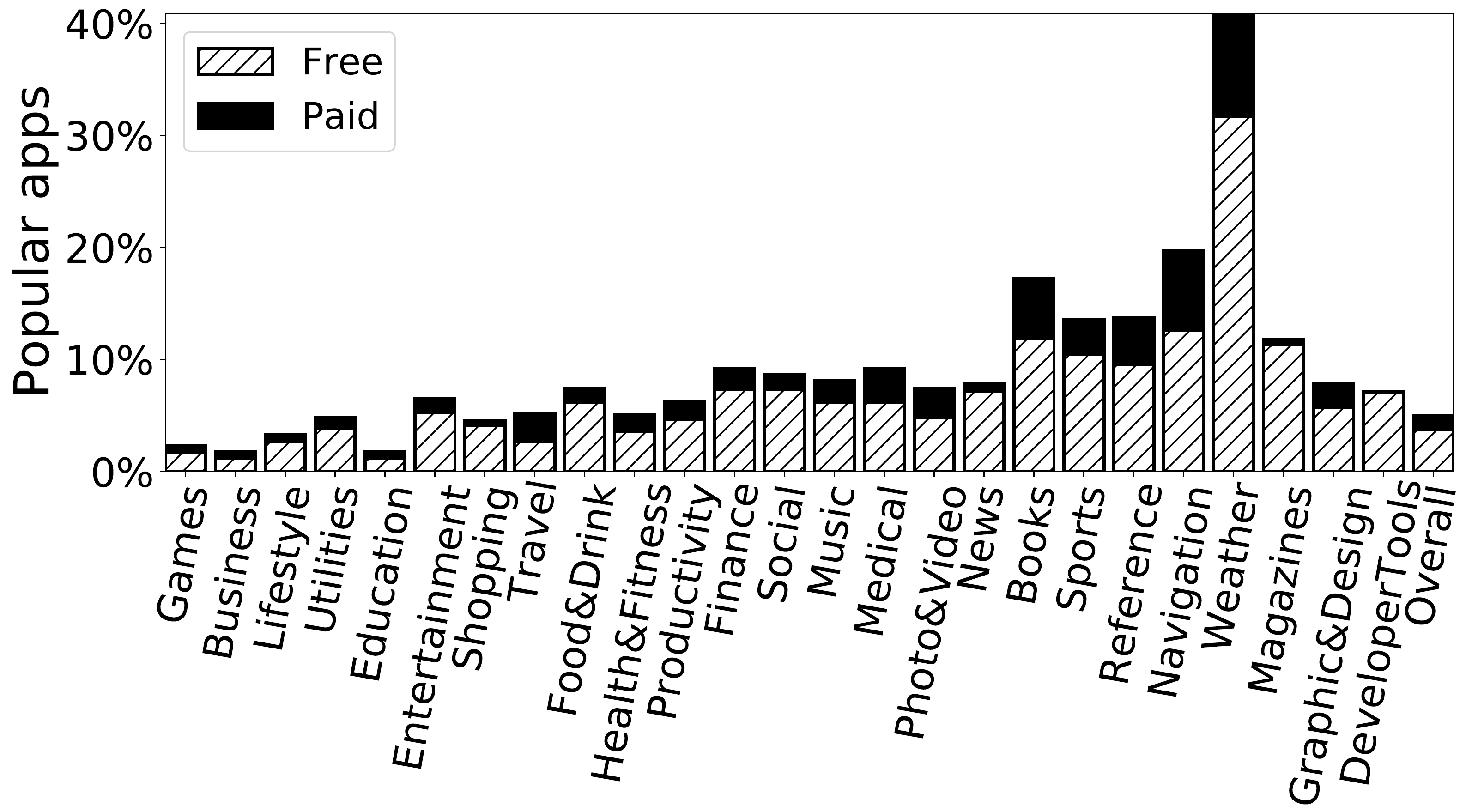}
  \caption{Percentage of removed popular apps in each category. The rightmost bar shows the overall proportion.}
  \label{fig:category-popular}
\end{figure}

Figure~\ref{fig:category-popular} presents the proportion of removed popular apps in each category. Overall, 51,704 of the removed apps (5\%) has been ranked in top-1500 across categories, while the remaining 95\% of removed apps have never got into the app ranking list during their lifecycle. 
\textit{This observation is basically in line with our expectation that most of the removed apps are non-popular apps.}
However, subtle differences arise when looking in detail on the per-category basis. For the removed apps, over 10\% of them are popular apps in categories including \texttt{Finance}, \texttt{Books}, \texttt{Sports}, \texttt{Reference}, \texttt{Navigation}, \texttt{Weather}, and \texttt{Magazines \& Newspapers}. One extreme category is \texttt{Weather}, and over 40\% of the removed apps have ever ranked in top-1500. 
In contrast, the removed popular apps only take up roughly 1.8\% to 2.3\% for categories including \texttt{Business}, \texttt{Education} and \texttt{Games}. 
There are two major reasons to explain this scenario. On one hand, app categories with higher percentage of removed popular apps are generally small categories without many apps. On the other hand, our manually investigation suggests that some policy-violation apps (e.g., illegal gambling apps) usually hide themselves in some specific categories including \texttt{Sports}, \texttt{Weather} and \texttt{Reference} (see Section~\ref{sec:removedpopular}), which makes the proportion of removed popular apps higher in these categories.

Furthermore, we examine the distribution of free and paid removed popular apps in these categories.
It is no doubt that free apps take up the majority (75.7\%) of removed popular apps. Nevertheless, 12,541 of the removed popular apps are paid apps that have been downloaded many times. For example, some paid gaming apps (e.g., Supreme Glory - Eternal Blood, Sword in the Air and Sword Dynasty) are quite popular even rank top-10 before removal. However, they were found to exaggerate their functional descriptions and post fake positive reviews to entice users to download them for revenue.
\textit{It suggests that the low-quality and potentially harmful apps have caused serious impact to unsuspecting users.}

\subsubsection{The number of user ratings}
\begin{figure}[t]
    \centering
    \begin{center}
        \includegraphics[width=0.8\textwidth]{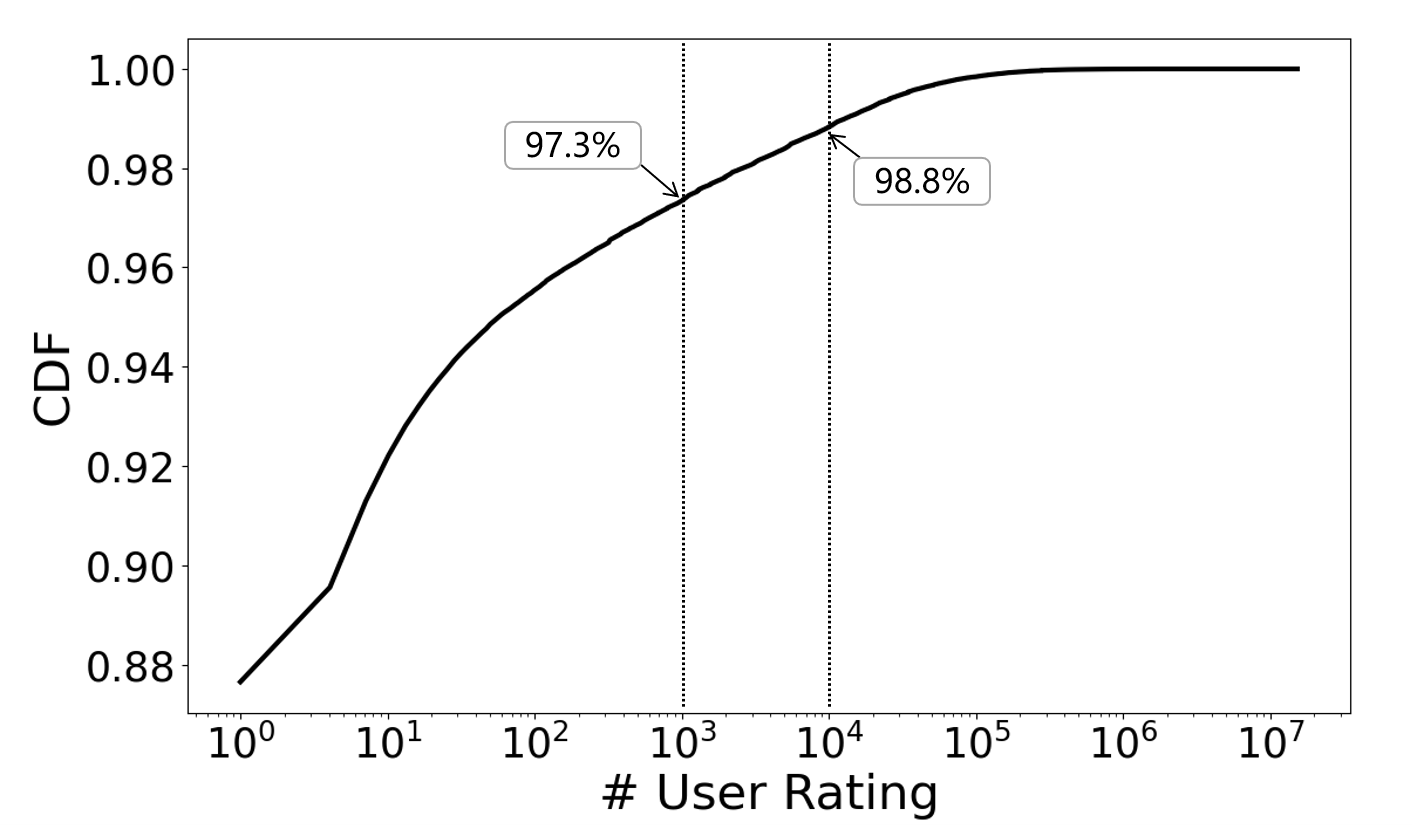}
        \caption{Distribution of the number of user rating.}
        \label{fig:rating-cdf}
    \end{center}
\end{figure}

The number of user ratings can be served as an implicit indicator of app popularity. In general, the more popular an app is, the more user ratings it will receive~\cite{iOSStoreStudy}. Figure~\ref{fig:rating-cdf} shows the overall distribution of the number of user ratings received by the removed apps. 
\textit{As expected, most of the removed apps (85.47\%) have received no user ratings, which further indicates that most of them are low-quality and abandoned apps with almost no active users.} 
Nevertheless, some removed apps have gained a considerable number of user ratings, i.e., roughly 2.7\% and 1.2\% of them have received more than 1,000 and 10,000 user ratings, respectively.  
For example, \texttt{Fawn weather pass}, one described as weather forecasting but actually a gambling app, has received 119,759 user ratings by the time of its removal.
\textit{It suggests that these apps have a large number of active users at least during some period of their life-cycle}. We will further explore the underlying removal reasons in Section~\ref{sec:removedpopular}.

\subsection{Developers of Removed Apps}

As aforementioned, 420,955 developers have contributed to the over 1 million removed apps in our dataset.
We next investigate these developers from two perspectives: 1) \textit{are there any developers who tend to release policy-violation apps?} It can be perceived by analyzing the proportion of removed apps in all the apps they released; and 2) \textit{are the removed apps dominated by a small number of aggressive developers?} It can be perceived by analyzing the developers with the most number of removed apps.

\begin{figure} [htbp]
\centering
\subfigure[CDF of developers by the percentage of removed apps]{
\label{fig:developer-1}
  \includegraphics[width=0.48\textwidth]{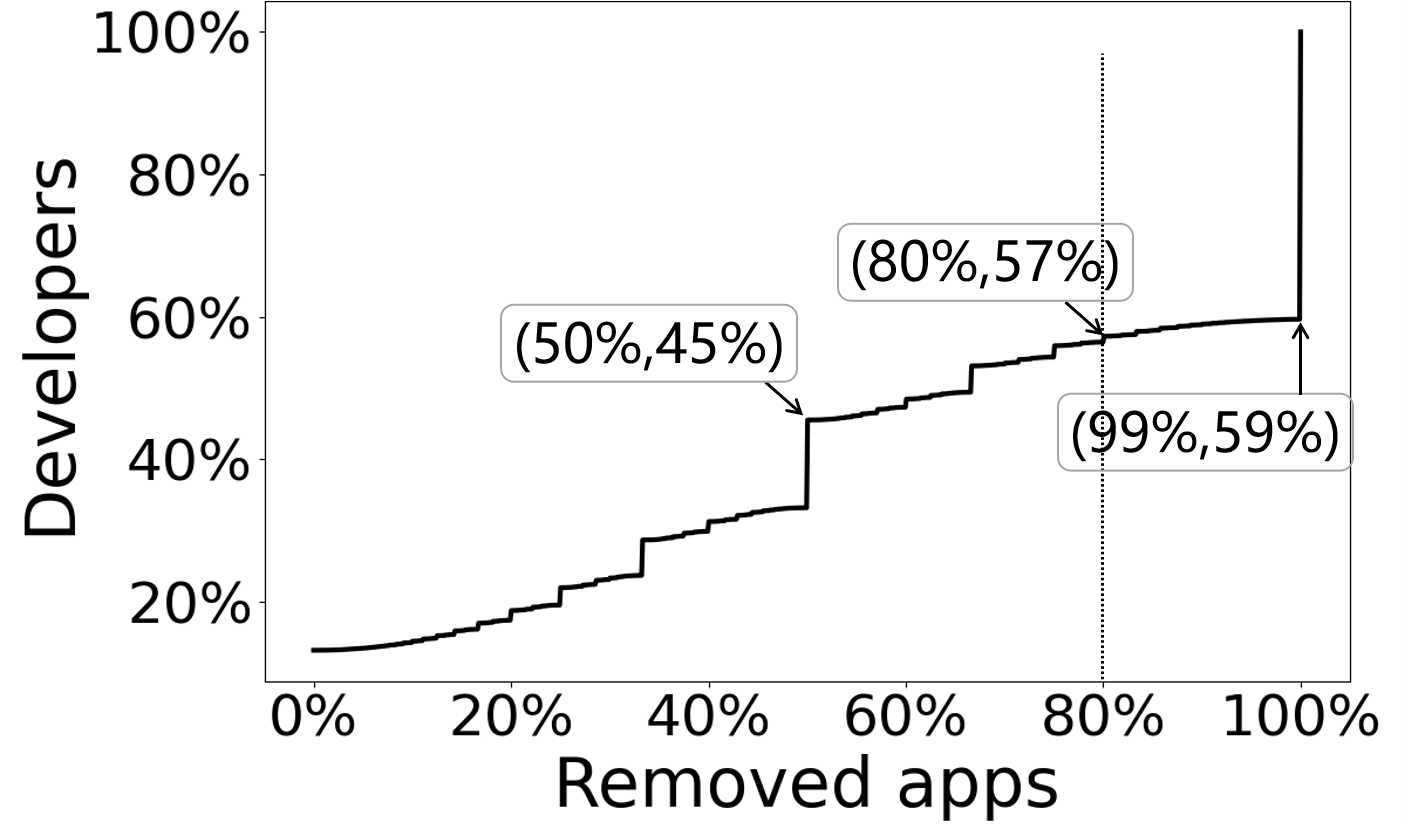}}
\subfigure[CDF of removed apps released by top developers]{
\label{fig:developer-2}
  \includegraphics[width=0.48\textwidth]{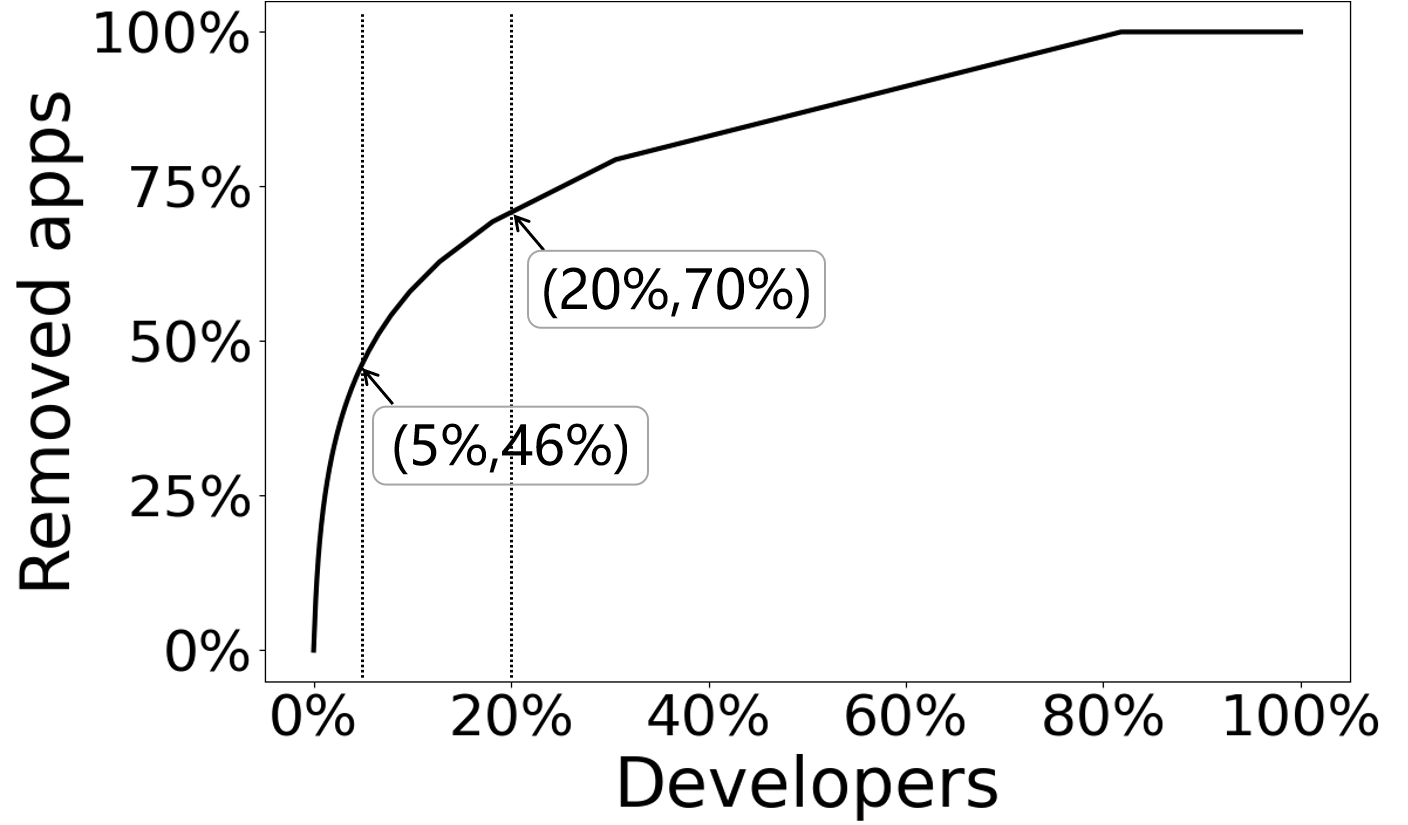}} 
\caption{The  developers of removed apps.}
\label{fig:developers}
\end{figure}

\subsubsection{Proportion of removed apps per developer.} 
We have collected the detailed app release history for all the developers we considered, through which we can get the proportion of removed apps for each developer.
Figure~\ref{fig:developer-1} presents the overall distribution. Note that, over 230K developers (55\%) in our dataset have released only one app and that app was removed, thus the proportion of removed apps for each of them is 100\%. We eliminate them in Figure~\ref{fig:developer-1} to make the distribution more clear. 
It can be seen that more than half of developers have over 55\% of their apps being removed from the app store. More seriously, 43\% of developers have more than 80\% of their apps being removed, 41\% of developers even have all their apps removed. 
When further considering that 55\% of developers have released only one app and that app was removed, 73.45\% of developers in our dataset have all their apps removed. 
\textit{It indicates that some developers tend to release policy-violation apps that would be removed by market. Thus, the app market and app users should pay special attention to the apps released by such kind of developers.}

\subsubsection{The most aggressive developers.}
We rank developers by the number of their removed apps, and observe that the maximum number of removed apps per developer is 100, with 438 such developers in total. Figure~\ref{fig:developer-2} shows the distribution of the number of removed apps per developer sorted in descending order. Interestingly, it follows the typical \textit{power law distribution}. Top 5\% of developers have contributed to over 46\% of the removed apps, and top 20\% of developers have contributed to over 70\% of the removed apps.
\textit{This observation suggests that majority of removed apps are dominated by a limited number of developers. We argue that the app market should design specific mechanism to label the trustworthiness of each developer, and raise alarm to the apps released by developers with aggressive behaviors.}

\begin{figure}[htb]
    \centering
    \begin{center}
        \includegraphics[width=0.8\textwidth]{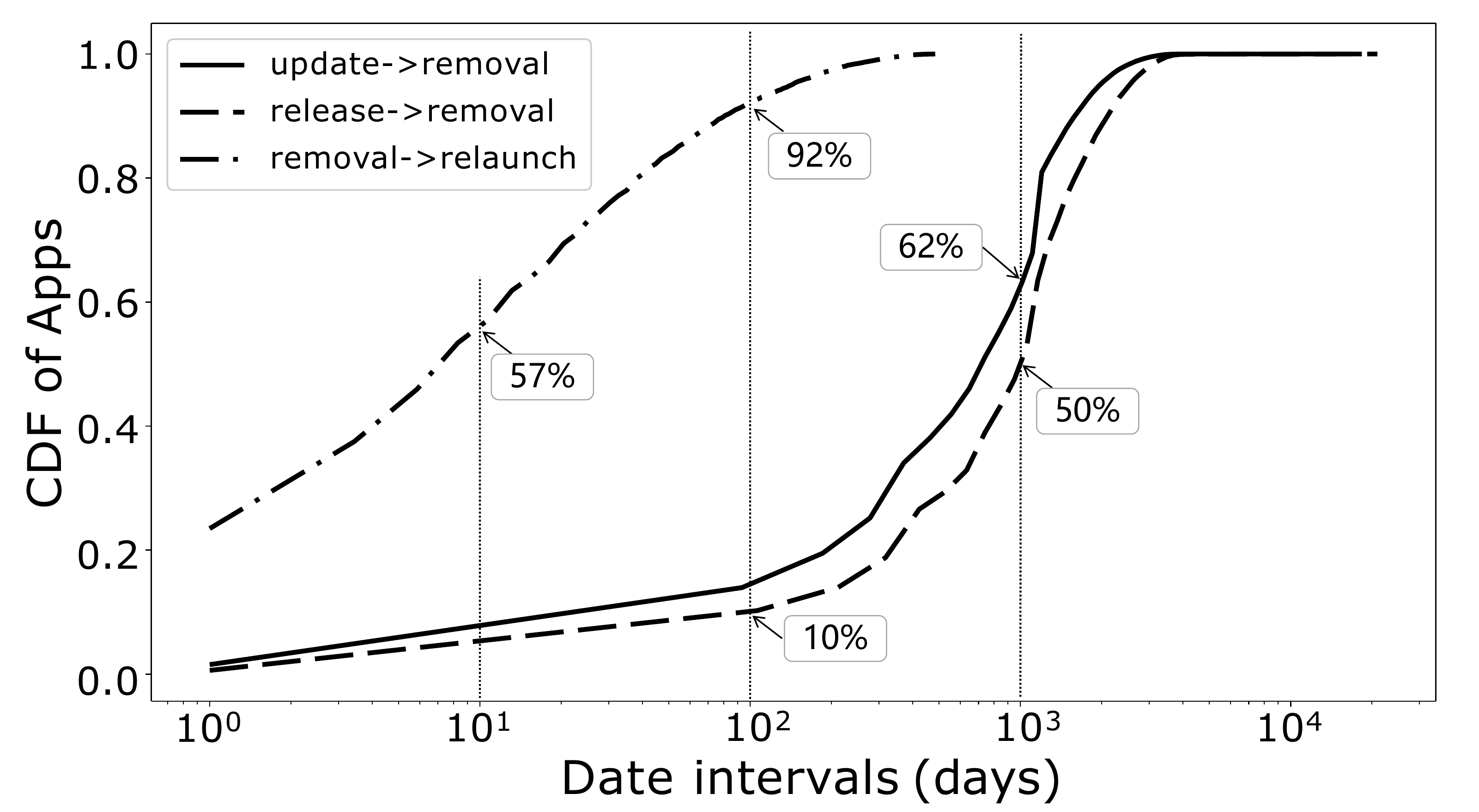}
        \caption{The distribution of time interval between update date and removal date, release date and removal date, removal date and relaunch date.}
        \label{fig:date-interval}
    \end{center}
\end{figure}

\subsection{Life-cycle of Removed Apps} 
\label{subsec:lifecycle}

\subsubsection{Life-cycle Analysis}
We next analyze the life-cycle of removed apps, i.e., from their launch (release date), to their updates (update date), to their removal (offline date). Note that, the removed apps can be re-launched to the market later, thus we record their re-launch date as well for further analysis.

\textbf{Update Date VS. Removal Date}
Figure~\ref{fig:date-interval} presents the time interval between the \textit{offline date} and the \textit{update date} of each removed app. 
A long interval of time reflects that it is quite possible the app has been outdated or abandoned by the app developer. 
Roughly 67\% of removed apps were not updated within 1 year before being removed, and even over 38\% of them remain silent within 1,000 days before being removed. \textit{It suggests that a large portion of the removed apps are outdated ones that may cannot function as intended in new version systems or no longer meet current app store guidelines.}

\textbf{Release Date VS. Removal Date}
We further pay attention to the time interval between \textit{app release date} and \textit{removal date}, which is the overall lifespan of that app. 
This metric can reveal the exposure time of a removed app, i.e., \textit{how long is it visible to mobile users?}
As shown in Figure~\ref{fig:date-interval}, only 10\% of the removed apps have a lifespan within 100 days and half of them last longer than 1,000 days. \textit{This observation indicates that majority of the removed apps can sustain in the app store for long time.} The longer the time is, the more negative effects it will introduce.

\textbf{Removal Date VS. Re-launch Date} 
As aforementioned, some of removed apps would be relaunched after a period of time. In our dataset, there are a total of 362,926 (35.1\%) removed apps were then relaunched (once or multiple times). To measure how long it typically takes for a removed app to be relaunched after being removed, we calculated the distribution of the date interval between each removal and relaunch, as shown in Figure~\ref{fig:date-interval}. The smaller the time interval, the sooner the app was re-launched. \textit{This metric can be used to measure developer's responsiveness to the controversial apps.}
Of all the relaunched apps, 8.7\% are re-launched within two day as they were removed and about 57\% are re-launched within 10 days after removal.
This observation suggests that \textit{most developers tend to fix the issues and relaunch their apps in a rapid manner}.

\begin{framed}
\noindent \textbf{Answer to RQ1:} 
\textit{
During the span of 1.5 years we studied, over 1 million apps were removed from iOS app market, which is surprisingly higher than our expectation. App removal in iOS app store shows some cyclical pattern. 
Although most of the removed apps are non-popular apps, 5\% of them are popular apps with high rank in app store and have gained a considerable number of user ratings. Some developers tend to release controversial apps that were removed, and the majority of removed apps are dominated by a limited number of developers. Most removed apps can sustain in the app store for long time.
}
\end{framed}

\section{Demystifying Removed Popular Apps}
\label{sec:removedpopular}

Our previous exploration suggests that, although most of the removed apps are low-quality apps (i.e., outdated and abandoned apps with almost no users), a number of the removed apps are quite popular (see Figure~\ref{fig:category-popular}). 
Over 5.0\% of them were ever ranked in top-1500 of their corresponding categories, and even 0.75\% of them were ever ranked in top-100 when competing with other apps in the same category. This observation motivates us to investigate the practical reasons behind the removal of such popular apps.

\subsection{Labelling the Benchmark}

\textbf{Candidate Selection.}
To demystify the removed popular apps, we first rely on manual effort to label some apps as the benchmark. To be specific, we select the removed apps whose most recent app rankings are within top-10 across each category before being removed from the market. In our dataset, there are 854 such apps in total, we believe they are definitely to be popular apps according to their app ranking.
To understand their characteristics, we further select normal popular apps that ranked in top-10 as comparison, with 759 such apps in total. Note that, as aforementioned in Section~\ref{sec:background}, we have gathered all their available information. 

\noindent \textbf{Manually Labelling.}
To figure out the reasons why apps are removed from the app store, the first two authors try to label the reasons according to the review guideline of iOS app store (see Section~\ref{sec:background}) independently. 
Both of the two authors have substantial mobile app analysis experience. To be specific, we rely on the detailed information we collected for labelling. 
For example, we will manually analyze all the user comments before app removal (including user complaints and potential fake reviews), investigate the variation of app popularity including their covered keywords, and check the app update records (e.g., some apps have substantially revised their app description and even developer information to cover some keywords that irrelevant to their contents), etc. Each removed app is labeled by two authors independently. If there exist divergence between two authors, the app will be extracted for further discussion to resolve disagreements. Finally, we figure out the practical reasons of app removal for 722 apps (84.5\%) out of the 854 apps. 
Note that for the remaining over 100 removed popular apps, we cannot find any clue from the information we collect to infer their removal reasons. It is quite possible that their app content has some controversial contents, or they were removed due to some political reasons (e.g., due to the censorship by some governments, iOS app store would also remove some popular apps.). We will further discuss it in Section~\ref{sec:discussion}. Nevertheless, our manual effort can successfully infer the mis-behaviors leading to app removal. 

Table~\ref{table:removal-reasons} presents the practical reasons leading to app removal for the 722 apps. They were classified into five major types.
Note that the removal reasons may not be mutually exclusive for a given app.
Most apps are removed mainly due to the reasons including \texttt{Ranking Fraud} (67.6\%), \texttt{Fake Description} (56.1\%), and \texttt{Content Issue} (55.4\%).
It is worth noting that there exist a large amount of gambling apps in our benchmark. These apps usually contain illegal contents, provide fake description (to evade the app vetting process), and disrupt app ranking using malicious ASO techniques (i.e., covering a large number of sensitive or irrelevant keywords).

Note that, during our manually labelling, we found almost all the mis-behaviors in these removed popular apps can be reflected in \textit{app review}, \textit{app description} and \textit{ASO keywords}. That is also why we can manually label their removal reasons as aforementioned.
For example, for the \texttt{Ranking Fraud} behavior, malicious actors always manipulate the app description and fake app reviews in an irregular manner to boost app ranking. Moreover, the ASO keywords it covered usually have quite different patterns with other normal apps in its category (see Figure~\ref{fig:keyword-number}). 
For the \texttt{Fake Description} behavior, the semantic information of its description is usually inconsistent with real users' comments (complaints) in app reviews.
For the apps with illegal content (\texttt{Content Issue}), they usually cover the corresponding illegal/sensitive keywords (e.g., gambling related keywords) during their app search optimization, in order to spread to as many users as possible in a short time.
For the remaining removal reasons, we found they are usually not isolated, while most of them would combine behaviors including \texttt{Ranking Fraud}, \texttt{Fake Description} or \texttt{Content Issue}.

Therefore, we will next detail the behavior patterns of removed popular apps from app review (Section~\ref{sec:pattern-review}), ASO keywords (Section~\ref{sec:pattern-aso}), and app description (Section~\ref{sec:pattern-description}), respectively. These characteristics would be used as indicators to facilitate our detection of these apps with mis-behaviors in Section~\ref{sec:detecting}.

\begin{table*}[]
\centering
\caption{The Manually Labelled Removal Reasons of Removed ``Popular'' Apps.}
\label{table:removal-reasons}
\begin{tabular}{lll}
\hline
Removal Reason & \# Apps & Description \\ \hline
Ranking Fraud & 488 & Apps that manipulate its metadata (including reviews) to boost its app ranking\\ 
Fake Description & 405 &  App description is inconsistent with app functions/behaviors\\ 
Content Issue & 400 & Apps that contain illegal content \\
In-app Purchase Fraud & 119 &  Apps that trick users into purchasing premium services \\
General Quality Issue & 105 &  Apps that have bugs, poor user experiences, crash issues, etc. \\
\hline
Overall & 722 & - \\
\hline
\end{tabular}
\end{table*}

\begin{table}[h!]
\centering
\caption{The number of app reviews collected.}
\label{table:review-info}
\begin{tabular}{|l|r|r|}
\hline
 & \multicolumn{1}{c|}{Removed Apps} & \multicolumn{1}{c|}{Normal Apps} \\ \hline
\# Reviews & 15,675,218 & 469,041 \\ \hline
\# Reviewers (Users) & 8,145,665 & 436,655 \\ \hline
\end{tabular}
\end{table}

\subsection{Behavior Patterns of App Review}
\label{sec:pattern-review}
App stores enable users to give feedback in the form of app reviews, which provides an opportunity to proactively collect user complaints and promptly improve apps' user experience. However,it is difficult to guarantee the quality and credibility of app reviews. During our exploration, we observe that the removed popular apps usually aggressively manipulate app reviews to boost their app ranking. This behavior is explicitly forbidden by iOS app store.
Next, we characterize the behavior patterns of these removed apps from three perspectives, including the \textit{the number of duplicated reviews} and \textit{the proportion of 5-star reviews} which are highly related to app ranking fraud, and the \textit{number of abnormal reviewers (users)} who posted the suspicious reviews.
The statistics of the app reviews we collected for the labelled 854 removed popular apps and the 759 normal popular apps is shown in Table~\ref{table:review-info}. Note that, we only collect their most recent 1 month reviews. Overall, there are over 16 million reviews posted by over 8 million unique app users in total. Obviously, the removed popular apps have much more user reviews (18,355 reviews on average, and 9,538 unique reviewers on average) than the normal apps (618 reviews on average and 575 unique reviewers on average).

\begin{figure} [htbp]
\centering
\subfigure[Duplicated reviews]{
\label{fig:duplicated-reviews-cdf}
  \includegraphics[width=0.48\textwidth]{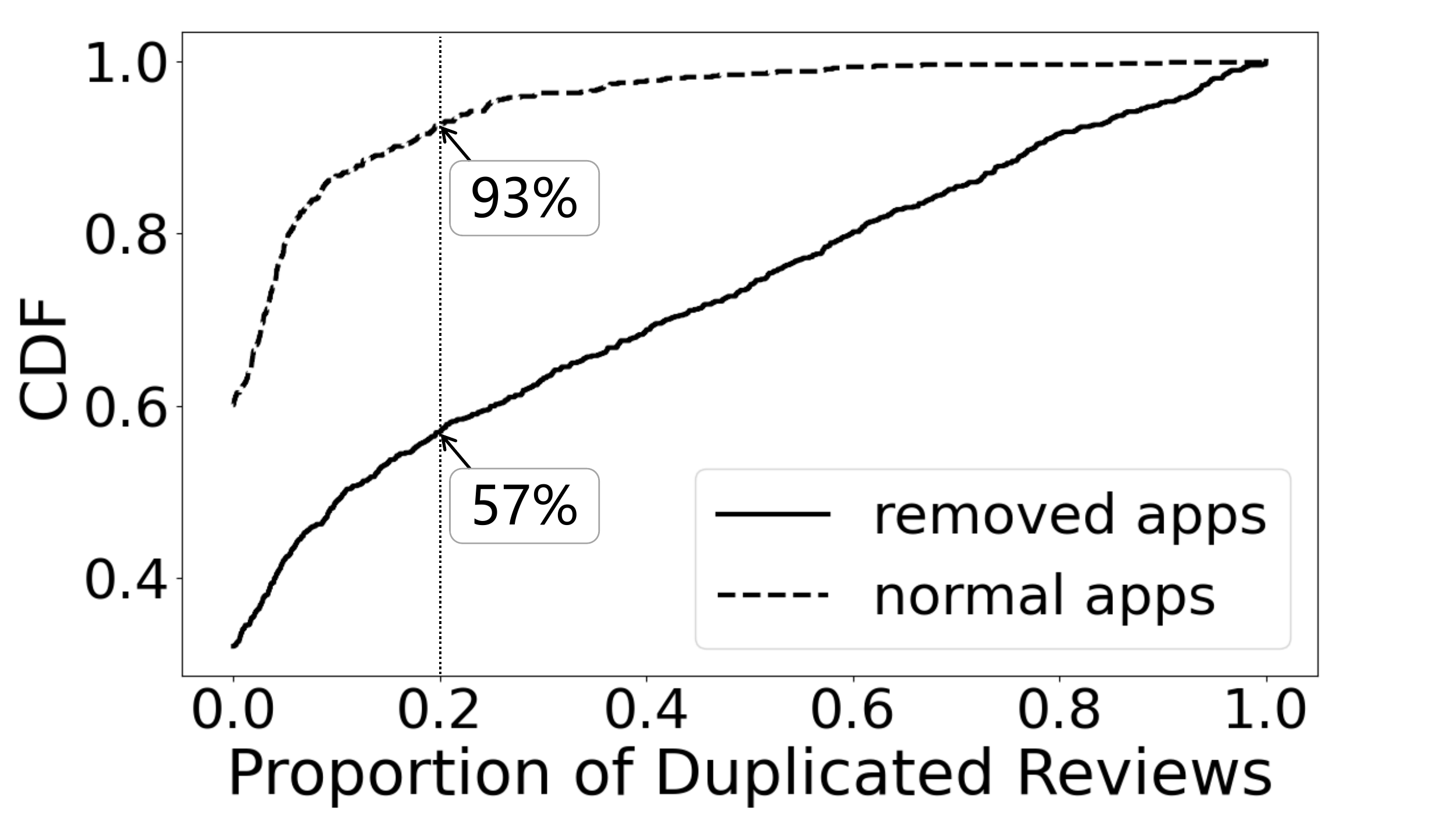}}
\subfigure[5 star reviews]{
\label{fig:5stars-reviews-cdf}
  \includegraphics[width=0.48\textwidth]{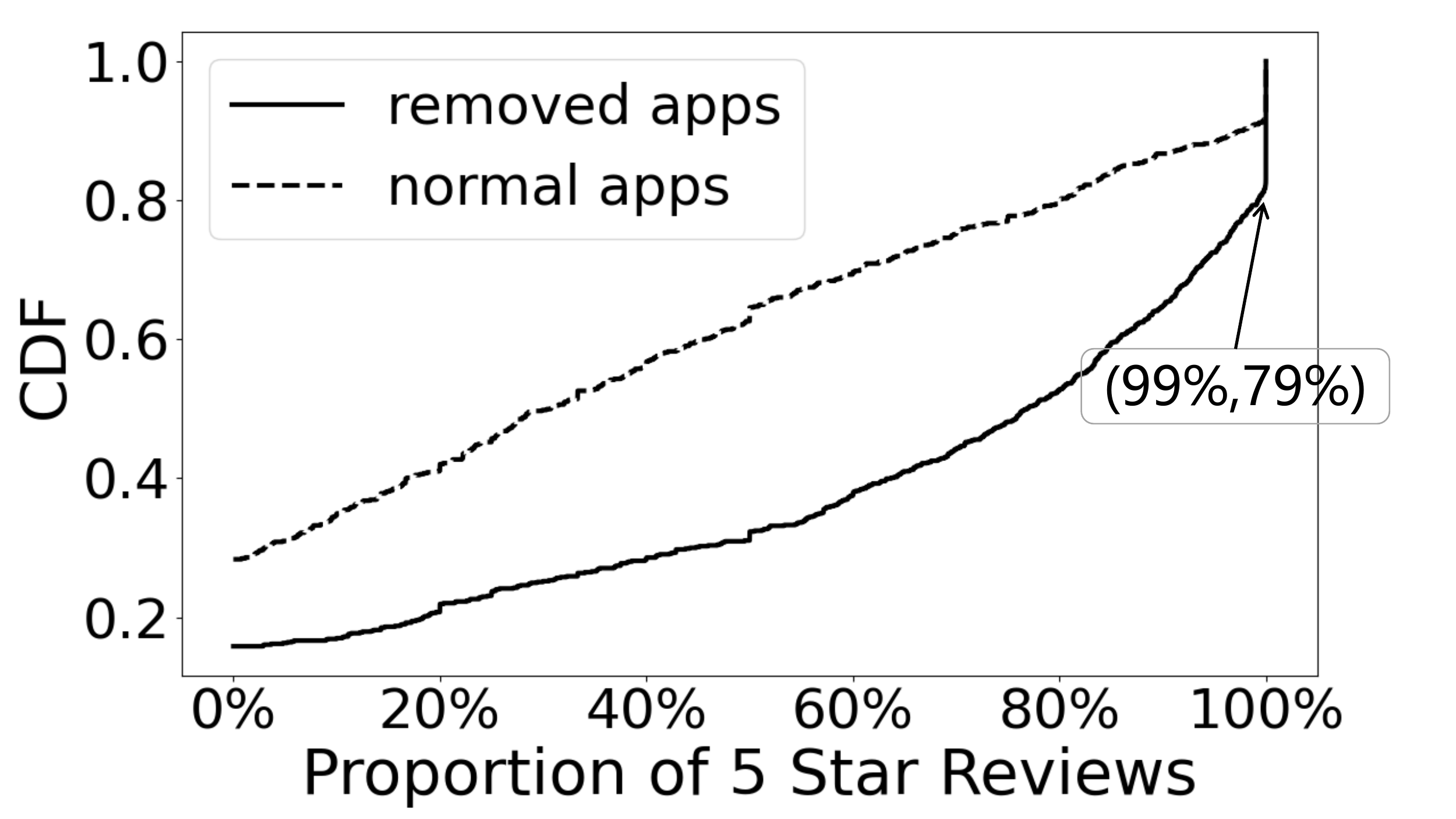}} 
\caption{The distribution of duplicated reviews and high score reviews in removed apps and normal apps}
\label{fig:reviews-cdf}
\end{figure}

\subsubsection{Duplicated Reviews}
Previous work~\cite{scambot,moneymaking} suggested that the duplication of app reviews can be a simple way to reflect the credibility of reviews.
If an app has a higher percentage of duplicated reviews, there is a higher possibility that the app has performed ranking fraud. 
In this work, we simply label a review is duplicated as long as there is another review posted within the app with identical content.
Figure~\ref{fig:duplicated-reviews-cdf} presents the proportion of duplicated reviews in both the removed popular apps and normal popular apps.
Obviously, the percentage of duplicated reviews in removed apps is much higher than that in normal apps.
Over 43\% of removed apps have more than 20\% of duplicated reviews, compared to only 7\% in normal apps. Over 8.5\% of the removed apps have more than 80\% of duplicated app reviews (VS. 0.4\% in normal apps). 
This result suggests that \textit{it is highly suspicious that there exist app ranking fraud in the reviews of the removed popular apps, and the proportion of duplicated reviews can be served as an indicator}.

\subsubsection{5-star Reviews}
In addition to text feedback, users can also review the app by giving a star rating, which is a number between 1 (bad) and 5 (good). 
We then characterize the proportion of high score reviews, i.e., 5-star reviews, of removed popular apps. 
Figure~\ref{fig:5stars-reviews-cdf} presents the distribution of 5-star reviews for both removed popular apps and normal popular apps. 
We can observe a higher proportion of 5 star reviews in removed apps than in normal apps. 
Surprisingly, there are over 20\% of removed apps whose proportion of 5 star reviews is close to 100\%. 
Considering that the removed apps are the ones which in general have some issues, the proportion of 5 star reviews for these apps is expected to be lower. 
By manually analyzing a large number of such reviews, we found that they are either irrelevant to app content (like advertisement) or duplicated reviews we identified. 
It further indicates that \textit{the app reviews in the removed apps are highly suspicious to be manipulated}.

\subsubsection{Abnormal Users}\label{subsubsec:abnormal-users}
The prior exploration suggests the high proportion of suspicious reviews. We next seek to explore the presence of abnormal users. Since different users are less likely to give the same review, users who give the same reviews are potentially to be abnormal users. 
In our analysis, reviews that have more than M words and appear N times in the dataset are considered as \textit{abnormal reviews}. 
Users who have given \textit{abnormal reviews} are considered as \textit{abnormal users}. 
To filter potential false positives, we flag abnormal reviews and abnormal users under two different conditions: 1) M equals to 5 and N equals to 10; and 2) M equals to 10 and N equals to 20.
Under the first condition, there are 15,350 abnormal users in the normal apps accounting for 3.58\% of all users who have commented the normal apps. In contrast, there are 5,099,175 abnormal users in the removed apps, which accounts for 62.6\% of all users who have commented the removed apps.  In the second condition, there are 15,187 abnormal users in normal apps (3.54\%), compared to 3,958,413 abnormal users in the removed apps (48.6\%).
The results suggest that \textit{a large portion of users in the removed popular apps are suspicious to be abnormal users, while the proportion of potential abnormal users is quite low in normal apps.} 

\subsection{Behavior Patterns of ASO Keywords}
\label{sec:pattern-aso}

As aforementioned, ASO is the process of improving the visibility of the mobile app. The goal is to make the app highly ranked in the search results of some keywords. iOS app store disallows developers to use excessive and misleading keywords. Our prior manual investigation suggests that the ASO keywords have been extensively manipulated by app developers. 

\begin{figure} [htbp]
\centering
\subfigure[Motivating Example]{
\label{fig:keyword-number}
  \includegraphics[width=0.48\textwidth]{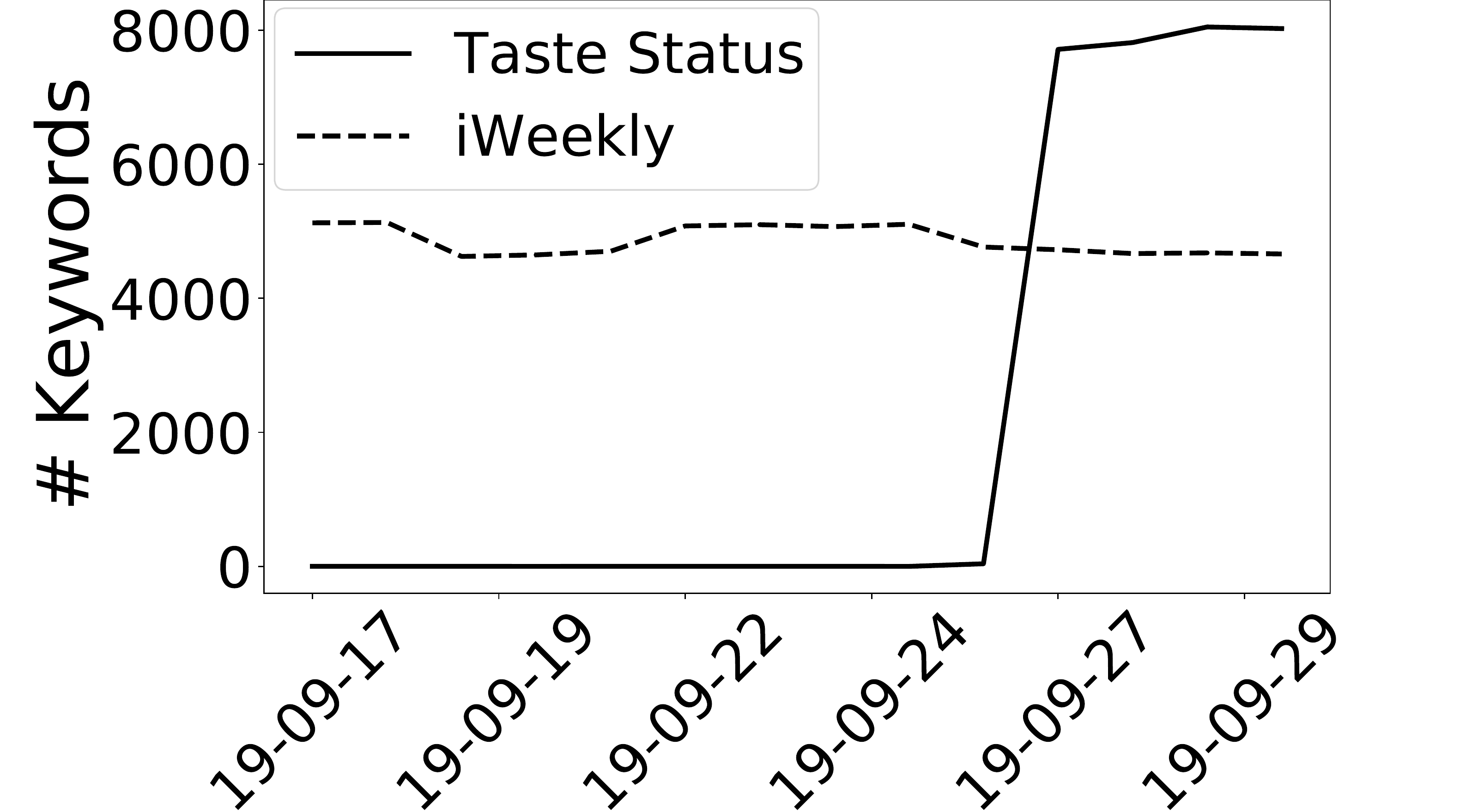}}
\subfigure[Standard Deviation]{
\label{fig:keywords-number-deviation}
  \includegraphics[width=0.48\textwidth]{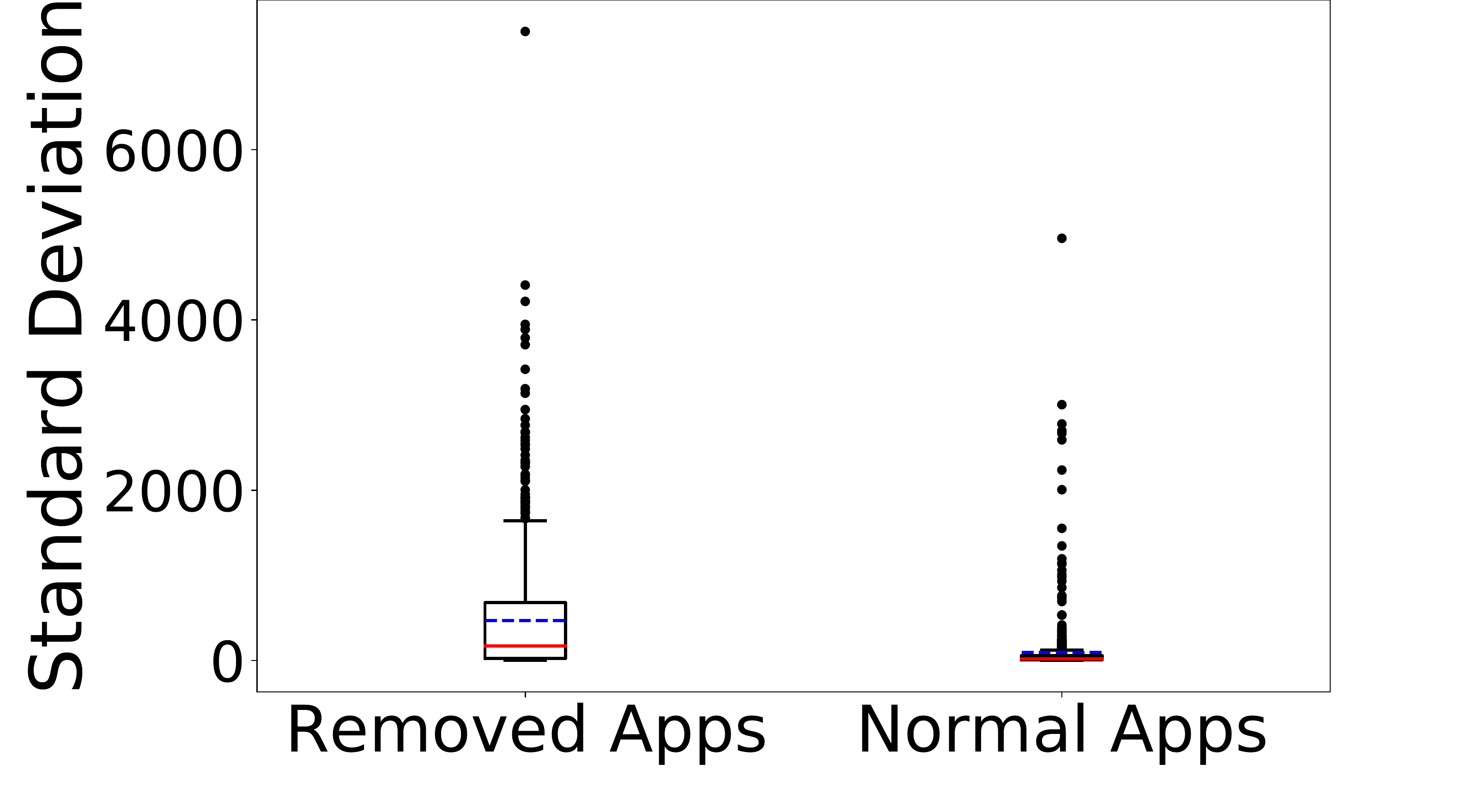}} 
  \vspace{-0.15in}
\caption{Behavior patterns of ASO keywords.}
\vspace{-0.2in}
\label{fig:keywords}
\end{figure}

\textbf{Motivating Example.}
We first start by giving a motivating example of the evolution of ASO keywords for a normal popular app and a removed popular app, respectively. Figure~\ref{fig:keyword-number} presents the variation of the number of daily keywords between 17th September, 2019 and 30th September, 2019. \texttt{Taste Status} is a gambling app which is removed from the app store on 1st October, 2019, while \texttt{iWeekly} is a normal popular app which belongs to \texttt{Magazines \& Newspapers} category. It can be observed that the number of keywords of Taste Status increased sharply during the period. The number of keywords was less than 10 at the very beginning. However, it increased to more than 7,500 within a single day. By manually analyzing the newly added keywords, we found most of them are irrelevant to its functionalities. 
In contrast, the number of keywords of \texttt{iWeekly} kept stable during the time period. It varies between 4,500 and 5,200. 
This example indicates that the characteristics of ASO keywords are quite different between normal apps and removed apps.

\textbf{The Standard Deviation of ASO Keywords}
\label{subsubsec:keyword-number}
Our initial exploration suggests the the variation of the numbers of keywords of normal apps and removed apps are quite distinct. 
Thus, we calculate the variation of the number of keywords of each app. 
Interestingly, there are 300 out of 722 (41.6\%) removed apps whose ASO keywords number increased by 1,000 within the week before being removed, while there are only 26 out of 759 (3.4\%) normal apps whose keywords number increase by 1,000. This indicates that the number of keywords of removed apps is more likely to increase sharply, while that of normal apps tends to keep stable. 
Figure~\ref{fig:keywords-number-deviation} presents the distribution of standard deviation of the number of ASO keywords. It can be observed that the standard deviation of the number of ASO keywords in removed apps are greater than that of normal apps. It further supports that the number of ASO keywords in removed apps in general changes rapidly.

\subsection{Behavior Patterns of App Description}
\label{sec:pattern-description}

\textbf{App Description VS. ASO Keywords}
\begin{table}[]
\centering
\caption{Keywords Coverage by Description of Removed Apps and Normal Apps}
\label{table:keyword-coverage}
\begin{tabular}{|c|r|r|}
\hline
\multicolumn{1}{|l|}{} & \multicolumn{1}{c|}{Removed Apps} & \multicolumn{1}{c|}{Normal Apps} \\ \hline
Keyword Coverage (avg) & 11.56 & 54.65 \\ \hline
App \# (Coverage Rate = 0) & \multicolumn{1}{r|}{97 (11.36\%)} & \multicolumn{1}{r|}{9 (1.05\%)} \\ \hline
\end{tabular}
\end{table}
Since there are sufficient keywords for a single app, we are interested in the content of keywords. It is expected that the contents of keywords are relevant to the app itself. Since the description introduces the main function of an app, the keywords should be covered by the descriptions. Therefore, we investigate the number of keywords covered by the description of apps. We first filter the apps whose average number of keywords is too small (i.e., the average number of keywords is smaller than 100). Then we calculate the average keyword coverage by description of removed apps and normal apps. Table~\ref{table:keyword-coverage} presents results. The average keyword coverage of removed apps is 11.56, while that of normal apps is 54.65. Over 11\% of removed apps have an average keyword coverage rate equal to 0, compared to only 1.05\% for normal apps. \textit{This observation suggests that the removed apps may cover many irrelevant keywords in the process of their app search optimization.}

\begin{framed}
\noindent \textbf{Answer to RQ2:} 
\textit{
The removed ``popular'' apps are mainly due to reasons including ranking fraud, fake description, content issues, in-app purchase fraud and general quality issues. Interestingly, by analyzing the meta information only, we can figure out the removal reasons for most of the removed popular apps. Some strong indicators from the app meta information can be used to differentiate removed apps from normal apps.
}
\end{framed}
\section{Detecting Removed Apps}
\label{sec:detecting}

Our prior exploration suggests that the indicators extracted from the app meta information (app description, app review, and ASO keywords) can be used to differentiate the apps that should be removed from normal apps. It motivates us to design an automated approach to flag suspicious apps that should be removed. Therefore, in this section, we seek to explore whether we can identify the controversial apps based on meta information only without accessing their app binaries (see Section~\ref{subsec:prediction}). More importantly, as it may take rather long time for the app store to remove potential harmful apps, which may cause negative impacts on users, thus we are wondering if we can further flag the policy-violation apps in advance (see Section~\ref{subsec:advance}).

\subsection{Predicting the Removed Apps}
\label{subsec:prediction}

We first perform a binary classification on the labeled apps in Section~\ref{sec:removedpopular}, to predict the apps that should be removed by combining all the indicators we identified from app meta information.

\begin{table}[t]
\centering
\caption{Features used for App Removal Prediction.}
\label{table:feature}
\resizebox{\linewidth}{!}{

\begin{tabular}{l|l}
\hline
Feature Name & Description \\ \hline
Review Count & The number of daily reviews. \\ 
Review Standard Deviation & The standard deviation of daily review number. \\ 
Rating Percentage & The proportion of reviews of different ratings. \\ 
Duplicate Reviews Percentage & The proportion of duplicated reviews. \\ 
Abnormal Users & The number of users posting duplicated reviews. \\ \hline
ASO Keyword Count & The number of daily keywords. \\ 
ASO Keyword Standard Deviation & The standard deviation of daily keyword number. \\ 
\hline
\# Keyword-Description Coverage & The number of keywords covered by the description \\ 
\% Keyword-Description Coverage & The percentage of keywords covered by the description. \\ \hline
\end{tabular}
}
\end{table}

\textbf{Feature Selection.}
Table~\ref{table:feature} shows the features used in the prediction. 
According to the behavior patterns we summarized in Section~\ref{sec:removedpopular}, we apply three sets of features. 
The first set of features represents characteristics of app reviews. These include the variation of reviews (number of daily reviews, standard deviation of the number of daily reviews), and the ratings of reviews (the proportion of reviews with different ratings). We also extract features from the suspicious reviews, including the proportion of duplicated reviews, and the number of users who have posted duplicated reviews.
The second set of features is extracted from ASO keywords, including the number of daily keywords and standard deviation of the number of daily keywords.
The third set of features is extracted from app description, including the number and the proportion of ASO keywords covered by app description.

\textbf{Machine learning models.}
We select five representative models for the classification tasks, which are Logistic Regression, Support Vector Machine (a.k.a, SVM), K-Nearest Neighbors (a.k.a, KNN), Random Forest, and Gradient Boosting Decision Tree (a.k.a, GBDT).
The implementation is based on scikit-learn~\footnote{\url{https://scikit-learn.org/stable/}} and  LightGBM~\footnote{\url{https://lightgbm.readthedocs.io/en/latest/}}.

\textbf{Evaluation Metrics.}
We use the widely used metrics including area under curve (a.k.a, AUC) score, precision, recall, F1 score, and accuracy to evaluate the performance of our model. 
A larger AUC score, precision, recall, F1 score, and accuracy indicates a better model. We perform a 10-fold cross validation and report the average performance on each of the metric, respectively.

\begin{table}[t]
\centering
\caption{Results of Prediction of App Removal.}
\label{table:result-full}
\begin{tabular}{|c|c|c|c|c|c|}
\hline
 & AUC & Precision & Recall & F1 & Accuracy \\ \hline
LR & 0.9028 & 0.8624 & 0.7778 & 0.8058 & 0.8180 \\ \hline
SVM & 0.9021 & 0.8909 & 0.7600 & 0.8056 & 0.8236 \\ \hline
KNN & 0.8703 & 0.8853 & 0.6725 & 0.7482 & 0.7874 \\ \hline
RF & 0.9000 & 0.8661 & 0.7918 & 0.8168 & 0.8328 \\ \hline
GBDT & 0.9116 & 0.8589 & 0.8214 & 0.8326 & 0.8397 \\ \hline
\end{tabular}
\end{table}

\textbf{Results.}
Table~\ref{table:result-full} presents overall result.
GBDT achieves the best performance in AUC score, recall, F1 score, and accuracy, while SVM achieves the best performance in precision. The best performances for AUC score, precision, recall, F1 score, and accuracy are 0.9116, 0.8909, 0.8214, 0.8326, and 0.8397, respectively. 
The results suggest that it is feasible to predict whether an app would be removed from the market based on its meta information.
We next investigate the feature importance in the classification. \textit{The number of abnormal users} is the most important feature while predicting the removal of an app. As indicated in Section~\ref{subsubsec:abnormal-users}, the proportion of abnormal users is much higher in removed apps. Besides, \textit{Keyword Coverage Percentage}, \textit{Keyword Standard Deviation}, and \textit{Review Standard Deviation} are also important while predicting removal apps. This is in line with our findings in Section~\ref{sec:removedpopular}

\subsection{Predicting the Removed Apps in Advance}
\label{subsec:advance}

To investigate whether we can flag the policy-violation apps in advance, we conduct a series of attempts, i.e., predicting whether an app will be removed from one day in advance to six days in advance.
Therefore, we utilize the features extracted from a shorter time period. For example, when we predict whether an app will be removed a day in advance, we will exclude the information we obtained in the last day and extract the features from the remainder information. \textit{Our expectation is that the policy-violation apps may release some initial signals that can be caught by our model.}

\begin{figure}[htb]
    \centering
    \begin{center}
        \includegraphics[width=0.8\linewidth]{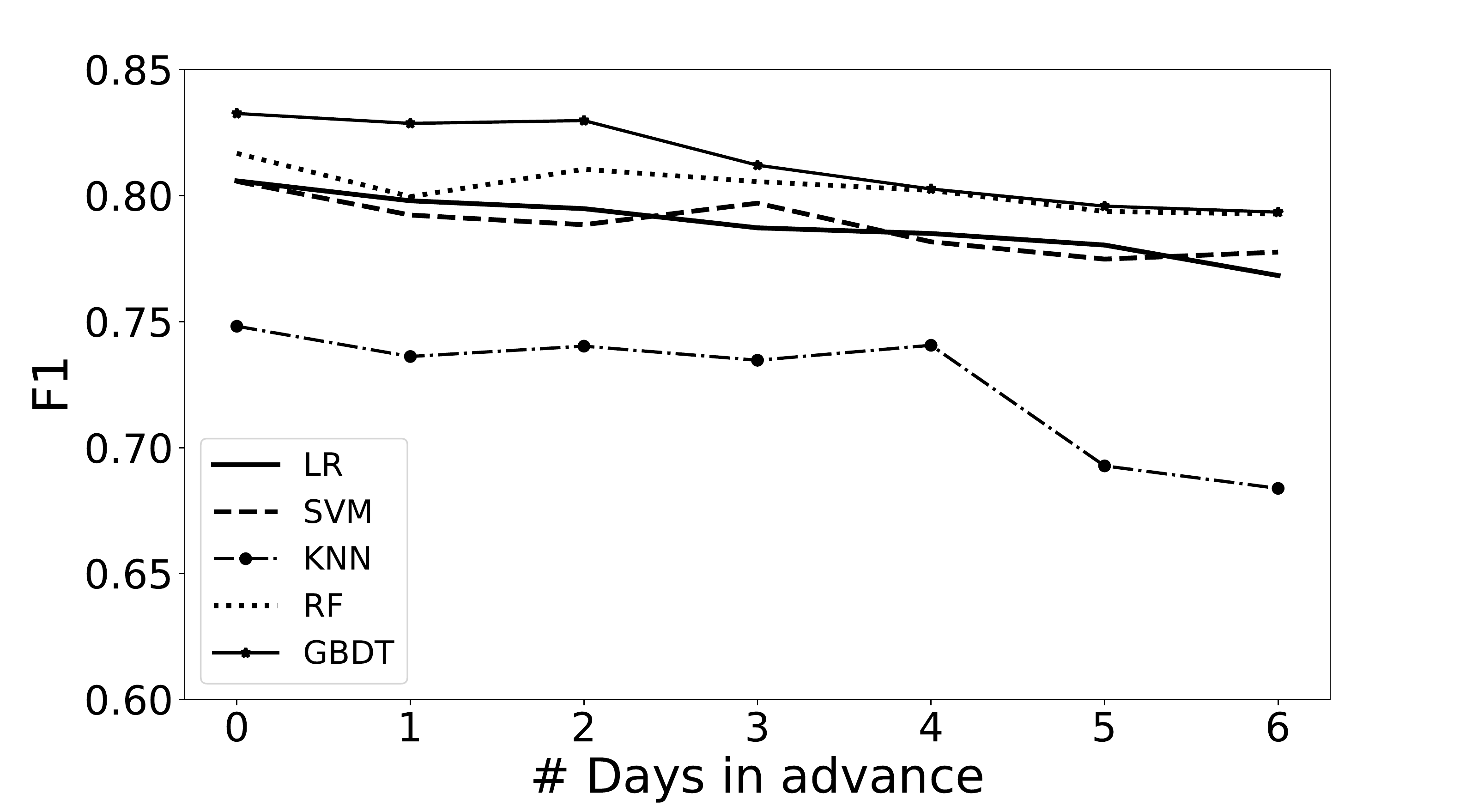}
        \caption{Predicting removed apps in advance.}
        \label{fig:advance-accuracy}
    \end{center}
\end{figure}

\textbf{Experiment Results.}
Figure~\ref{fig:advance-accuracy} presents the overall result (F1 score) of app removal prediction from 0 day to 6 days. 
Obviously, the overall performance drops slightly along with the increment of time in advance for all models. Nevertheless, we observe the prediction result is quite good. For example, even we only rely on the information obtained prior to 6 days, the best model (GBDT) can achieve an F1 score of 79.3\%. \textit{Therefore, this result suggests that it is feasible to advance the process to identify and remove the suspicious apps from the market.}

\begin{framed}
\noindent \textbf{Answer to RQ3:} 
\textit{
Experiment results suggest that, even without accessing to app binaries, we can identify the removed apps with good performance (F1=83\%) based on the indicators extracted from app meta information. Moreover, even with fewer information, we are able to accurately flag the removed apps 6 days in advance.
}
\end{framed}
\section{Discussion}\label{sec:discussion}
In this section, we discussed the findings across platforms (i.e., Android and iOS), the insights and implications for practice, and the limitations that can potentially affect the results of our study. 
\subsection{iOS App Store VS. Google Play}
Our study is focused on iOS app store, one of the universally acknowledged most secure app stores, due to its locked-down ecosystem. Our study reveals a number of issues in the ecosystem, which we believe to exist in Android app markets like Google Play as well. Similar spamming techniques abused by the removed apps can be directly adopted to Google Play without any technical barriers. Moreover, our study in this paper does not rely on app bytecode, thus our investigation methods can be transferred to analyzing Google Play apps easily. The only concern is to harvest the comprehensive dataset of apps in Google Play. Our investigation is built atop of the unique app removal dataset that was collected extensively by monitoring the iOS app market daily across the span of 1.5 years. To the best of our knowledge, there is no such kind of datasets available for Android app markets yet.

\subsection{Implications}
Our observations are of key importance to the stakeholders in the mobile app ecosystem. 
\textit{First, it shows the ineffectiveness of market regulation of iOS app store}. A number of apps with mis-behaviors were found in the market and then were removed, which usually takes a rather long time. This, however, can be improved through proactively monitoring the whole market using techniques like the one proposed in this paper. Our observation shows that some developers have the tendency to release policy-violation apps, while the market should pay special attention to apps released by these developers.
Furthermore, we observe that the app ranking and app searching recommendation mechanisms in iOS app store are seriously abused by spam apps, while indicates the emergency for detecting and regulating this kind of behaviors. 
\textit{Second, app developers should be aware of the iOS app store guidelines to prevent their apps from being removed from the market}. Our observation suggests that, over 30\% of removed apps were re-launched to the market after a specific time. Although most of them seek to fix issues in a rapid manner, it would greatly impact the reputation and the popularity of the app.
\textit{Third, even the ``popular'' apps are not trustworthy, as they can abuse a number of ASO techniques to disrupt the app market}. Thus, mobile users should pay special attention to the suspicious high ranking apps.

\subsection{Limitations}
Our study, however, carries several limitations. 
First, in this paper, we did not touch the bytecode of iOS apps, while some removal reasons can only be analyzed via the mobile apps. Nevertheless, we show that most of the removed popular apps can be identified proactively with features extracted from app metadata only. In the future, we can combine static or dynamic analysis of app binaries to get a deeper understanding of the removed apps.
Second, this paper relies on some manual efforts to label the removal reasons. Although we strictly follow the iOS app store guidelines to understand their removal reasons, it is quite possible that our labelling reasons are inconsistent with their actual reasons. 
Third, as aforementioned, app removal can be perform by both app developers and the app market. However, it is almost impossible for us to to infer whether the app was removed by iOS app market.

\section{Related Work}\label{sec:related}

\subsection{Removed App Analysis}
This is the first comprehensive study of app removal practice in iOS app store at scale, longitudinally, and across various dimensions. To the best of our knowledge, the most related work is a preliminary study of removed apps in Google Play~\cite{wang2018android}. However, their work is coarse-grained as they only created two snapshots of Google Play in 2015 and 2017, which can only be used to roughly know which app was removed. They cannot infer the accurate app removal time, and they cannot measure the overall landscape of removed apps as we did including the daily trend, app popularity, and app life-cycle, etc. More importantly, we have collected the most comprehensive dataset including app reviews and ASO keywords, which are not covered by their work. Based on such information, we can analyze the behaviors patterns of the removed apps and design an automated approach to detect them, which were not touched by existing studies.


\subsection{Mobile App Analysis}
Mobile app analysis has been widely explored from various aspects, such as malware detection~\cite{gorla2014checking, zhou2012dissecting, grace2012riskranker}, app privacy and security~\cite{almuhimedi2015your, tu2018your, welke2016differentiating, taylor2017update, iCryptoTracer,PiOS,ProtectMyPrivacy,chen2016following,xu2017appholmes}, app usage~\cite{li2015characterizing, zhao2016discovering, bohmer2011falling,li2016voting,liu2017understanding,DBLP:journals/tois/LiuALTHFM17,lu2017prado,DBLP:conf/icse/LuLLXMH0F16,DBLP:conf/www/LiLAMF15,li2020systematic}, and so on~\cite{lu2016learning,chen2020comprehensive,CHENDLDEPLOY2}.
These efforts focus on detecting and understanding the characteristics of malware, revealing the privacy and security problems of mobile apps, and characteristic of the usage pattern of different users under different situations. However, most of the existing studies are focused on Android app analysis, while only a few of them were focused iOS apps~\cite{iCryptoTracer,PiOS,ProtectMyPrivacy, brown2014100, khalid2013identifying, benenson2013android, chen2016following}. 
For example, iCryptoTracer~\cite{iCryptoTracer} was proposed to detect misuse of cryptography functions in iOS apps based on static and dynamic analyses.
Chen et al.~\cite{chen2016following} proposed to detect potentially-harmful libraries in iOS apps based on the observations that many iOS libraries have their Android versions that can potentially be used to understand their behaviors and relations.
PiOS~\cite{PiOS} was proposed to detect privacy leakage for iOS apps by analyzing data flows in Mach-0 binaries.
ProtectMyPrivacy~\cite{ProtectMyPrivacy} is a system designed to detect access to private data by iOs apps and protect users by substituting anonymized data. 
Existing methods which aim to identify malware and potential harmful apps are mostly based on code analysis, which can be served as a complementary of this work to understand the deeper reasons of app removal.

\subsection{App Ecosystem Analysis}

There are plenty of efforts focusing on analyzing mobile app ecosystem through the perspective of app marketplaces~\cite{petsas2013rise, wang2019understanding, wang2018beyond, wang2017explorative, chen2014ar, martin2016survey}.
For example, Petsas et al.~\cite{petsas2013rise} analyzed the mobile app ecosystem through four popular third-party Android app marketplaces. They found out that app downloads follow a power-law distribution. Wang et al.~\cite{wang2019understanding, wang2018beyond} analyzed the evolution of mobile app ecosystems through Google Play and Chinese Android app markets. They found that although the overall mobile app ecosystem shows promising progress, there still exists a number of unsolved issues for all mobile app marketplaces. The condition in Chinese app marketplaces is even worse. Chen et al.~\cite{chen2014ar} proposed AR-Miner, which aims to extract informative informative reviews from app marketplaces. 
These efforts mainly focus on analyzing the overall patterns of apps and developers in different app marketplaces. However, these studies are limited to investigate the characteristics of removed apps like this paper, as it usually needs a continuous effort to monitor the market.

\section{Conclusion}
\label{sec:conclusion}

In this paper, we have conducted a large-scale and longitudinal study to understand the removed apps in iOS app store. Specifically, our analysis covers over 1 million removed apps obtained by monitoring iOS app market daily across the span of 1.5 years. Overall, our analysis suggests that the number of removed apps is surprisingly high, and the removal behaviors show cyclical patterns in iOS app store. We further pay special attention to the removed popular apps and design an automated approach to predict whether an app would be removed. Experiment results suggest that we can accurately predict the removal of apps, even we are able to flag them several days in advance. We believe our research efforts can positively contribute to the mobile app ecosystem and promote best operational practices across app markets.

\balance
\bibliographystyle{ACM-Reference-Format}
\bibliography{sample-base}




\end{document}